\documentstyle[twoside,fleqn,espcrc2]{article}

\newcommand{\eqn}[1]{(\ref{#1})}
\newcommand{\ft}[2]{{\textstyle\frac{#1}{#2}}}
\def\be{\begin{equation}}
\def\ee{\end{equation}}
\def\bea{\begin{eqnarray}}
\def\eea{\end{eqnarray}}

\renewcommand{\a}{\alpha}

\newcommand{\pa}{\partial}
\newcommand{\g}{\gamma}

\renewcommand{\l}{\lambda}

\newcommand{\n}{\nu}

\newcommand{\s}{\sigma}

\def\IZ{{\hbox{{\rm Z}\kern-.4em\hbox{\rm Z}}}}

\def\bigone{{\hbox{1\kern -.23em{\rm l}}}}
\title{Supersymmetric quantum mechanics, supermembranes and 
Dirichlet particles }
\author{Bernard de Wit\address{Institute for Theoretical Physics,
Utrecht University\\
Princetonplein 5, 3508 TA Utrecht, The Netherlands }}
\begin{document}
\setlength{\arraycolsep}{0pt}
\begin{titlepage}
\begin{flushright} THU-97/02\\ hep-th/9701169
\end{flushright}
\vfill
\begin{center}
{\large\bf Supersymmetric quantum mechanics, supermembranes and 
Dirichlet particles${}^\dagger$}\\
\vskip 7.mm
{Bernard de Wit }\\
\vskip 0.1cm                                                      
{\em Institute for Theoretical Physics} \\
{\em Utrecht University}\\
{\em Princetonplein 5, 3508 TA Utrecht, The Netherlands} 
\end{center}
\vfill
\begin{center}
{\bf ABSTRACT}
\end{center}
\begin{quote}
We review models of supersymmetric quantum mechanics that are 
important in the description of supermembranes and of Dirichlet 
particles, which play a role in the context of M-theory.  
\vfill      \hrule width 5.cm
\vskip 2.mm
{\small\small
\noindent $^\dagger$ Invited talk given at the $30^{\rm th}$ 
International Symposium Ahrenshoop on the Theory of Elementary 
Particles, Buckow, August 27 - 31, 1996; to appear in Nuclear 
Physics B (Proc. Suppl.).}   
\end{quote}
\begin{flushleft}
January 1997
\end{flushleft}
\end{titlepage}
\begin{abstract}
We review models of supersymmetric quantum mechanics that are 
important in the description of supermembranes and of Dirichlet 
particles, which play a role in the context of M-theory.  
\end{abstract}
\maketitle
\setcounter{footnote}{0}
\section{Supersymmetric quantum mechanics}       
Consider the class of supersymmetric Hamiltonians
\bea
H&=& {\textstyle {1\over 2}} P_a^A\, P_{aA} + {\textstyle{1\over 4}} 
\Bigl( f_{ABC} X^B_a X^C_b\Bigr)^2 \nonumber \\
&& -{\textstyle {1\over 2}}i 
f_{ABC}\, \theta^A \gamma^aX_a^B\theta^C \,, \label{hamiltonian}
\eea
depending on a number of $d$-dimensional coordinates $\vec X^A$ 
($a=1,\ldots,d$) and corresponding 
momenta $\vec P^A$ as well as real 
spinorial anticommuting coordinates $\theta^A_\alpha$. The index 
$A$ labels the vectors and spinors as components of 
some (compact) Lie algebra $g$ with structure constants 
$f_{ABC}$. The phase space is restricted to an subspace invariant 
under the corresponding Lie group $G$, and is therefore subject 
to the constraints 
\be
\varphi_A =   f_{ABC}
\left( \vec X^B\!\cdot\!\vec P^C - {\textstyle{1\over 2}}i
\theta^B\theta^C\right) \approx 0\,. \label{constraint}
\ee
The above models were proposed long ago as extended models of 
supersymmetric quantum mechanics with more than four 
supersymmetries \cite{CH}. The spatial dimension $d$ and the 
corresponding spinor dimension is severely restricted. The models exist 
for $d=2,3,5$, or 9 dimensions; the (real) spinor dimension equals 
2, 4, 8, or 16, respectively. Naturally this is also the number of 
independent supercharges, defined by 
\be
Q=\left(P_a^A\gamma^a +{\textstyle{1\over 2}} f_{ABC} 
X^B_aX^C_b\,\gamma^{ab} \right) \theta^A \,.    \label{scharge}
\ee
These charges generate the familiar supersymmetry algebra (in the 
group-invariant subspace),
\be
\{Q_\alpha,Q_\beta\}\approx 2 H\, \delta_{\alpha\beta} \,
.\label{susyalg} 
\ee

At the time these models were proposed it was recognized that 
they coincide precisely with the zero-volume limit of 
supersymmetric gauge theories. Hence one suppresses the 
spatial dependence of   
the vector and spinor fields, while the time-like component 
of the gauge field is set to zero by a gauge choice:
\bea
A_\mu^A({\bf x}, t) \to \cases{ A_a^A(t)\,,\cr
\noalign{\vskip2mm}
A^A_0 = 0\,,  \cr} 
\qquad
\theta^A({\bf x}, t) \to  \theta^A(t)\,. 
\eea
The constraint equations (\ref{constraint}) then correspond to 
Gauss' law. These Yang-Mills theories are formulated in 
Minkowski spacetimes of dimension 3, 4, 6,  or 10.

Let us briefly mention a number of characteristic features that are 
extremely relevant for what follows. We note the 
appearance of  
``valleys'' where the potential in \eqn{hamiltonian} vanishes. 
This happens whenever the coordinates $\vec X^A$ are restricted 
to take values in an abelian  
subalgebra. These valleys extend all the way to infinity where 
they become increasingly narrow. Their existence raises questions 
about the nature of the spectrum of \eqn{hamiltonian}. In 
the bosonic versions of these models the 
wave function cannot freely extend to infinity, because at 
large distances it becomes more and more squeezed 
in the valley. By the uncertainty principle, this gives rise to 
kinetic-energy contributions which increase monotonically  
along the valley. Another way to see this 
is by noting that the oscillations perpendicular to the valleys 
give rise to a zero-point energy, which induces an effective 
potential barrier that confines the wave function. 
This confinement causes the spectrum to be discrete 
\cite{LuescherS}. However, for the supersymmetric  
models defined by \eqn{hamiltonian} the situation is different. 
Supersymmetry can cause a cancelation of the transverse 
zero-point energy. Then the wave function is no longer confined, 
indicating that the supersymmetric models have a 
continuous spectrum. The latter was proven in \cite{DWLN}.  
 
At the moment there is no proof that the Hamiltonian 
\eqn{hamiltonian} allows  normalizable or 
localizable zero-energy states, superimposed on the continuous 
spectrum. The relevance of such states will be discussed later 
on. There is an important difference 
between states whose energy is exactly equal to zero and states 
of positive energy. Because of the supersymmetry algebra, 
zero-energy states must be annihilated by the supercharges. 
Hence, they are supersinglets. These singlets may have some 
degeneracy, for instance, associated with some other symmetries 
such as rotational invariance. The positive-energy states, on the 
other hand, must constitute full supermultiplets. So they are 
multiplets consisting of multiples of  $1+1$, $2+2$, $8+8$, or 
$128+128$ bosonic $+$ fermionic states, corresponding to $d=2,3,5$ 
or 9, respectively.

\section{Supermembranes}                         
Fundamental supermembranes can be described in terms of actions 
of the Green-Schwarz type, possibly in a nontrivial 
(super)spacetime background \cite{BST,BTD}. Such
actions exist 
for supersymmetric $p$-branes, where $p= 0$, 1, $\ldots$ defines the
spatial dimension of the brane. Thus for
$p=0$ we have a superparticle, for $p=1$ a superstring, for
$p=2$ a supermembrane, and so on. The
dimension of spacetime in which the superbranes can live is
very restricted \cite{AETW}. These restrictions arise from the 
fact that the action contains a 
Wess-Zumino-Witten term, whose supersymmetry depends sensitively
on the spacetime dimension. If the coefficient of this term
takes a particular value then the action has an additional
fermionic gauge symmetry, the so-called $\kappa$-symmetry.
This symmetry is necessary to
ensure the matching of (physical) bosonic and fermionic
degrees of freedom.
In the following we restrict ourselves to 
supermembranes (i.e., $p=2$) in eleven dimensions.  
 
In the notation and conventions of \cite{DWHN} the
supermembrane Lagrangian reads
\bea
&&{\cal L} =- \,\sqrt{-g(X,\theta)} - \epsilon^{ijk} \,
\overline\theta\Gamma_{\mu\nu}\partial_k\theta \\ 
&&\;\times \Big [{\textstyle{1\over 2}}
\,\partial_i X^\mu (\partial_j X^\nu +
\overline\theta\Gamma^\nu\partial_j \theta) + {\textstyle{1\over 6}}
\,\overline\theta\Gamma^\mu\partial_i\theta\;
\overline\theta\Gamma^\nu\partial_j\theta \Big] \,, \nonumber 
\label{mblagrangian}
\eea
where $X^\mu(\zeta)$ and $\theta(\zeta)$ denote the superspace
coordinates of the membrane parametrized in terms of world-volume
parameters $\zeta^i$ ($i = 0$, $1$, $2$). The fermionic coordinates
$\theta$ are 32-component Majorana spinors. The gamma matrices 
are denoted by $\Gamma^\mu$; gamma matrices 
with more than one index denote antisymmetrized products of gamma
matrices in the usual fashion. The metric
$g_{ij}(X,\theta)$ is the induced metric on the world volume,
\be
g_{ij} = (\partial_iX^\mu + \overline\theta\Gamma^\mu
\partial_i\theta) (\partial_jX^\nu +
\overline\theta\Gamma^\nu
\partial_j\theta) \,\eta_{\mu\nu}\,, 
\ee
where $\eta_{\mu\nu}$ is the eleven-dimensional Minkowski 
metric. It is easy to see that $g_{ij}$, and therefore the first term in 
\eqn{mblagrangian},
is invariant under spacetime
supersymmetry. In $4$, $5$, $7$, or $11$ spacetime dimensions the 
second term proportional to $\epsilon^{ijk}$ is also 
supersymmetric (up to a total divergence) and the full action is 
invariant under  $\kappa$-symmetry.

In order to study the quantum-mechanical properties of the 
supermembrane we use the light-cone gauge. First we define 
light-cone coordinates 
$$
X^\mu = \cases{X^\pm = \displaystyle{1\over {\sqrt 2}}(X^{10} 
\pm X^0) \cr
\noalign{\vskip2mm}
X^a\, \qquad (a=1,\ldots,9 )\cr} 
$$ 
with a similar decomposition of the gamma matrices $\Gamma^\mu$ 
into $\gamma^\pm$ and $\gamma^a$ ($a=1,\ldots,9 $). 
It is possible to impose the gauge conditions 
\cite{DWHN,Goldstone,Hoppe,BSTa} 
\bea
 X^+ (\zeta) &=& X^+ (0) + \tau \,, \phantom{1\over x}\nonumber\\
\partial_r X^- &=& -{1\over P^+}\left(\vec P\cdot 
\partial_r\vec X 
+P^+ \overline\theta \gamma_- \partial_r \theta\right)\,,
\nonumber  \\
P^+ &=& P^+_0 \;\sqrt{w(\sigma)} \;,\qquad \gamma_+ \theta =  0\,
, \label{gauge} 
\eea
where $\tau=\zeta^0$ and the remaining two parameters $\sigma^r=  
(\zeta^1, \zeta^2)$ 
parametrize the spatial extension of the membrane. 
We have introduced conjugate momenta and a density 
$\sqrt{w(\sigma)}$ normalized according to $\int {\rm d}^2\! 
\sigma \; \sqrt{w(\sigma)} = 1$. Therefore the constant $P^+_0$ 
represents the  membrane centre-of-mass (CM) momentum in the 
direction associated with the coordinate $X^-$,
\be
P^+_0 = \int {\rm d}^2\! \sigma \,P^+ . 
\ee
The other CM coordinates and momenta are    
\bea
&&\vec P_0 = \int {\rm d}^2\! \sigma \; \vec P \,,  \qquad \vec 
X_0 = \int \!{\rm d}^2\!\sigma\sqrt{w(\sigma)}\, X(\sigma)\,, 
\nonumber \\ 
&& \theta_0 = \int \!{\rm d}^2\!\sigma 
\sqrt{w(\sigma)}\, \theta(\sigma)\,.  
\eea
In the light-cone gauge we are left with the transverse
coordinates $\vec X$ and corresponding momenta $\vec P$, which
transform as vectors under the $SO(9)$ group of transverse
rotations. Only
sixteen fermionic components $\theta$ remain, which transform as
$SO(9)$ spinors. Furthermore we have the CM momentum
$P_0^+$ and the CM coordinate $X^-_0$ (the
remaining modes in $X^-$ are dependent), while the CM momentum
$P_0^-$ is equal to minus the supermembrane Hamiltonian.
 
The supermembrane Hamiltonian takes the following form
\be
H = {{\vec P}_0^{\,2}\over 2 P_0^+} + {{\cal M}^2\over 2P_0^+}  
\,, \label{mbhamiltonian}
\ee 
where $\cal M$ is the supermembrane mass operator, which does {\it
not} depend on any of the CM coordinates or momenta. After some 
changes in notation of the spinors, the expression for ${\cal M}^2$ 
is 
\bea
&&{\cal M}^2 =  \int {\rm d}^2\!\s \; \sqrt{w(\s)}\;   \label{mass}\\ 
&&\;\times \Big[ {[\vec 
P^2(\s)]' \over w(\s)}  +\ft12 \Big(\{X^a,X^b\}\Big)^2 - i 
\theta\g_a\{\theta,X^a\}\Big]\,, \nonumber
\eea
where $[\vec P^2]'$ indicates that the contribution of the CM 
momentum $\vec P_0$ is suppressed. We made use of a Lie bracket 
$\{A, B\}$, which, for two functions $A(\sigma)$ and $B(\sigma)$, 
is defined by  
\be
\{A,B\}(\sigma)\equiv {1 \over \sqrt{w(\sigma)}} \,
\epsilon^{rs} \partial_r A(\sigma) \,\partial_s B(\sigma)\,. 
\label{bracket}
\ee
It is straightforward to show that \eqn{bracket} satisfies the Jacobi
identity. We stress that we assume that the membrane embedding 
coordinates are single-valued functions so that the contributions 
of the CM coordinates cancel also in the second and third term 
under the integral in \eqn{mass}. 

The mass operator is accompanied by a constraint, which follows 
from the integrability of the second gauge condition \eqn{gauge} 
for $\pa_rX^-$, 
\be
\varphi= \{X^a,P^a\} -\ft12 i\{\theta_\a ,\theta_\a\} 
\approx 0\,.\label{mbconstraint}
\ee
Apart from its CM value, $X^-$ can then be 
determined in terms of the other coordinates and momenta. 
         
The Hamiltonian commutes with 32 supercharges. Half of them 
are proportional to the fermionic CM spinor 
coordinates $\theta_0$. Therefore they have no bearing on the mass 
operator. The other 16 charges take the form 
\be
Q = \int {\rm d}^2\!\s \;\Big[ P^a \g_a +\ft12  \sqrt{w} \{X^a,
X^b\}\g_{ab}\Big] \,\theta \,.
\ee
These charges involve both the CM momenta $\vec P_0$ and the modes 
contained in the mass operator \eqn{mass}. They  satisfy
the supersymmetry algebra, in the constrained subspace,
\be
\left\{Q_\alpha ,
Q_\beta\right\}\approx \delta_{\alpha\beta}\, {\cal M}^2 \,,
\ee
 
The structure of the Hamiltonian \eqn{mbhamiltonian} shows that the wave
functions for the supermembrane now factorize into a
wave function pertaining to the CM modes and a wave
function of the supersymmetric quantum-mechanical system that
describes the other modes. For the latter the mass operator plays 
the role of the Hamiltonian. When the mass operator vanishes on the state, 
then the 32 supercharges act exclusively on the CM coordinates 
and generate a massless supermultiplet of eleven-dimensional 
supersymmetry. In case there is no other degeneracy beyond that 
caused by supersymmetry, the resulting supermultiplet is the one 
of supergravity, describing the graviton, the antisymmetric tensor 
and the gravitino. In terms of the $SO(9)$ helicity 
representations, it  
consists of ${\bf 44} \oplus {\bf 84}$ bosonic and $\bf 128$
fermionic states.
When the mass operator does not vanish on the states, we are 
dealing with huge supermultiplets consisting of multiples of 
$2^{15}+2^{15}$ states.

 
\section{Supermembrane as models in supersymmetric quantum 
mechanics}
The gauge conditions \eqn{gauge} adopted above leave a residual
invariance consisting of so-called area-preserving 
diffeomorphisms. They are defined by $\sigma^r 
\to \sigma^r + \xi^r(\sigma)$ with
\be
\partial_r\left(\sqrt{w(\sigma)} \,\xi^r(\sigma)\right) = 0\,. 
\label{areapres}
\ee
The general solution of \eqn{areapres} can be decomposed into
co-exact and harmonic vector fields. For a membrane of genus $g$
there are precisely $2g$ independent
harmonic vectors. Furthermore,
there can be homotopically nontrivial transformations; these
will not be
considered in what follows. The co-exact components are
parametrized in terms of globally defined functions $\xi(\sigma)$,
\be
\xi^r(\sigma) =  {{\epsilon^{rs}}\over \sqrt{w(\sigma)} }\,
\partial_s\xi(\sigma)\,,
\ee
and generate an {\it invariant} subgroup, which we denote by $G$.
In the following, we restrict our attention
to this invariant subgroup when refering to
area-preserving transformations.
The commutator of two infinitesimal $G$-transforma\-tions
characterized by functions $\xi_1$ and $\xi_2$ yields a third 
$G$-transformation characterized by $\xi_3$, which equals  
\be
\xi_3 = \{\xi_2, \xi_1\}\,,  \label{areaalg}
\ee
According to \eqn{areaalg} the bracket \eqn{bracket} 
thus encodes the structure constants of the group $G$.
This interpretation shows that the supermembrane mass 
operator \eqn{mass} belongs to the class of supersymmetric 
Hamiltonians \eqn{hamiltonian}. The constraint equation 
\eqn{mbconstraint} restricts the physical states to be invariant 
under the area-preserving diffeomorphisms. Clearly the 
supermembrane in a $(d+2)$-dimensional spacetime can be  
regarded as a limiting case of the supersymmetric quantum 
mechanical models in $d$ dimensions discussed in section~1. 

In order to elucidate this relation one expands all 
coordinates and momenta into a complete
orthonormal basis of functions consisting of the constant
function 1 and functions $Y^A(\sigma)$ (where $A= 1,
2,\ldots, \infty$). The coefficient of 1 represents the CM value,
so that we write
\be
\vec X(\sigma) = \vec X_0 + \sum_{A=1}^{\infty}
\,\vec X_A\, Y^A(\sigma)\,, 
\ee
with similar expansions for the other coordinates and momenta. 
The basis functions 
$Y^A$, which can be chosen real or complex, are normalized
according to
\bea
&&\int \!{\rm d}^2 \sigma \sqrt{w(\sigma)}  \;Y^A(\sigma)  =0\,,
\\
&&\int \!{\rm d}^2 \sigma \sqrt{w(\sigma)}  \;Y^A(\sigma) \,
Y_B(\sigma) = \delta^A_B \,, \nonumber
\eea
where indices $A$, $B$ are raised and lowered by complex 
conjugation. Their completeness implies  
\be
\sum_{A=1}^\infty \,Y^A(\sigma )\, Y_A(\sigma^\prime) =  
{1 \over \sqrt{w(\sigma)} }\,\delta^{(2)} (\sigma,
\sigma^\prime)-1\,.  
\ee
The bracket $\{Y^A, Y^B\}$ can be expressed in terms of the 
$Y^A$ and one derives 
\be
\{ Y^A, Y^B\} = f^{AB}{}_{\!\!C}\, Y^C\,,  
\ee
so that the constants $f^{AB}{}_{\!\!C}$ represent the
structure constants of the infinite-dimen\-sional group $G$. Other
tensors related to the diffeomorphisms generated by harmonic
vectors and tensors needed for the Lorentz algebra generators
were defined in \cite{DWMN}. 

If we replace $G$ by a finite group, then (13) defines the
Hamiltonian of a supersymmetric quantum-mechanical system.
In the limit to the infinite-dimensional group $G$ we thus recover the
supermembrane. This observation enables one to
regularize the supermembrane in a
supersymmetric way by considering a limiting procedure based on a
sequence of groups whose limit yields the group $G$. For closed 
membranes of certain topology it is known how to approximate the 
group $G$ as a particular $N\to \infty$ limit of $SU(N)$. To 
be precise, it can be shown that the structure constants of 
$SU(N)$ tends to those of $G$ up to corrections of order 
$1/N^2$. The nature of this limit is subtle and depends 
on the membrane topology. As long as $N$ is finite, no 
distinction can be made with regard to the topology. In some 
sense, all topologies are thus included at the level of finite 
$N$. Another subtlety concerns the emergence of 
diffeomorphisms associated with the harmonic vectors, which 
cannot be incorporated for finite $N$, at least not in 
infinitesimal form. For a discussion of these subtleties we 
refer to \cite{DWMN,DWN}.  

The classical configurations of zero energy are now characterized 
by $\{X_a,X_b\}=0$. This implies that the membrane collapses 
into stringlike configurations of zero area. The same feature 
exists for general $p$-branes. Classical (super)$p$-branes are unstable:
the zero-energy
configurations correspond to collapsed branes of lower
dimensionality $p-1$. Whether or not this unstability persists at 
the quantum level is the subject of the next section.
Of all $p$-branes, only the
(super)particle and the (super)string do not suffer from the 
above instability.
The particle simply because it has no internal structure at all,
and the string because, apart from its centre-of-mass motion, all
modes are confined by harmonic-oscillator potentials.

\section{Continuous spectrum}
As we pointed out above supermembranes and the 
supersymmetric quantum-mechanical models introduced in section~1 
share the important feature that the potential vanishes in 
certain valleys that extend to infinity.  
In the  presence of these valleys,
there is a latent danger that the wave functions will no
longer be confined. In that case there is then no obvious
reason why the spectrum should be discrete. However, as already 
discussed in section~1, 
quantum-mechanical effects may still prevent the wave function
from escaping through the zero-energy valley. To
make this more explicit, consider the following two-dimensional
Hamiltonian,
\be
H_B = p_x^{\,2} + p_y^{\,2} + x^2y^2\,. \label{HB}
\ee
Obviously the potential in \eqn{HB} has zero-energy valleys along 
the $x$- and the $y$-axis. Nevertheless the eigenfunctions of  
\eqn{HB} are confined and cannot escape through these valleys. 
This follows from decomposing \eqn{HB}  as
\be
H_B= \ft12 (p_x^{\,2}+ p_y^{\,2}) + H_1 +H_2, 
\ee
where $H_1 = \ft12 p_x^{\,2} + {1\over 2}x^2y^2$ and $H_2 = \ft12 
p_y^{\,2} +
{1\over 2}x^2y^2$. Since $H_1$ and $H_2$ take the form of
harmonic oscillator Hamiltonians in $x$ and $y$, respectively,
with frequencies equal to $|x|$ or $|y|$, we immediately
derive the operator inequality ($\hbar=1$)
\be
H_B \geq \ft12(p_x^{\,2}+ p_y^{\,2}))+ \ft12(|x|+|y|)\,. 
\ee
The operator on the right-hand side has a discrete spectrum as 
the wave function will be confined by the infinitely rising 
potential. Therefore  the spectrum of $H_B$ is discrete 
\cite{LuescherS}. 
 
It is now obvious why the introduction of supersymmetry could
drastically change the situation described above. As is
well-known, supersymmetric harmonic oscillators have no
zero-point energy, so that the confining effective potential may
vanish. Whether or not the potential in the valley will vanish
completely can be investigated in a simple two-dimensional model 
\cite{DWLN}. We first introduce a supersymmetry charge by
\be
Q=Q^\dagger = \s_x\,p_x +\s_y\,p_y +  \s_z\,xy \,,  
\label{toycharge}
\ee
where $\s_x$, $\s_y$, $\s_z$ are the Pauli spin matrices, so that 
$Q$ acts in a two-dimensional fermionic space. The 
Hamiltonian then follows in the usual fashion,
\be
H= Q^2 = \pmatrix{ H_B  & x+iy\cr
         \noalign{\vskip6mm}
          x-iy  & H_B  \cr}.   \label{toyhamiltonian}
\ee
 
In order to establish the absence of an effective
potential barrier that may prevent the wave function
from escaping through the valley, we consider a set of normalized 
trial wave functions
\be
\psi_\lambda (x,y) = \chi(x-\lambda) \; \varphi_0(x,y)\;\xi_{\rm 
F}\,, \label{wavefunction}
\ee
characterized by some parameter $\lambda$. Here $\chi$ is a
one-dimensional free-particle wave packet
satisfying $\int {\rm d}x\, |\chi|^2
=1$, which has compact support so that \eqn{wavefunction} is only 
different from zero for $x\approx \lambda$, $\varphi_0$ is the 
normalized groundstate wave function for a one-dimensional 
harmonic oscillator, 
\be
\varphi_0(x,y) = \pi^{-{1\over 4}}\, |x|^{1\over 4}\,
\exp {\left(-{\textstyle{1\over 2}}|x| \,y^2\right)}\,,
\ee
and $\xi_{\rm F}$ is a normalized two-dimensional spinor.
When the parameter $\vert\lambda\vert$
is large, the wave function \eqn{wavefunction} thus has its 
support in a narrow region along the $x$-axis. Irrespective of 
the precise form of $\chi(x-\l)$, we have the identity
\be
Q\,\psi_\l = \Big(\s_x\,p_x + xy\,\s_z ({\rm sgn}\,x\, \s_x  
+ {\bf 1} )\Big)\psi_\l \,. 
\ee
By suitably choosing the spinor $\xi_{\rm F}$ the second 
term vanishes, so that 
\be 
Q\,\psi_\l = \s_x\,p_x\psi_\l =- {\rm sgn}\,x\; p_x\,\psi_\l\,.
\ee 
One can easily verify that for this choice of  
$\xi_{\rm F}$ the zero-point energy in $H_B$ associated with the 
transverse oscillations (i.e., in $y$) cancels against the 
fermionic zero-point energy induced by the term $x\,\s_x$ in the 
Hamiltonian \eqn{toyhamiltonian}.  
 
What we now intend to show is that, by making $\vert\lambda\vert$ 
large, thus pushing the domain of support of the wave function 
deeper into the valley, $\psi_\lambda$ will tend to a 
supersymmetric wave function, 
which, by virtue of the supersymmetry algebra, must have zero
energy. To see how this works, we derive the following result for 
any finite power $\n$ of $Q$ acting on $\psi_\l$,
\be
\parallel Q^\n\,\psi_\lambda\parallel^2 = |(p_x)^\n\chi|^2 + {\rm
O}(\lambda^{-1}), \label{Qasympt}
\ee
where the first term on the right-hand side is obvious. It 
represents the norm of the 
one-dimensional wave function $(p_x)^\n\,\chi$, which is equal to the
$\n$-th power of the energy of the wave packet since $\chi$ is
normalized to unity. The second term originates
from operators $p_x$ acting on $\varphi_0$. This
introduces factors $|x|^{-1}$ or $y^2$; the latter
become also proportional to $|x|^{-1}$ upon 
integrating over $y$. As $x\approx \lambda$, the contributions
from $p_x\,\varphi_0$ are thus of order $\lambda^{-1}$.
 
Obviously the right-hand side of \eqn{Qasympt} can be made
arbitrarily small by choosing a wave packet $\chi$ of
sufficiently low energy and by making $\lambda$ sufficiently
large. The latter implies that the wave function will
extend further and further
into the valley, so that there is apparently no
confining force, as $\psi_\lambda$ approaches a supersymmetric
wave function with zero energy. Furthermore we derive along the 
same lines, for any $E$,
\be
\parallel (H-E)\psi_\lambda\parallel^2 = |(p_x^{\,2}-E)\chi|^2 +  
{\rm O}(\lambda^{-1})\,.\label{Hasympt}
\ee
For any positive $E$ and $\varepsilon$ we can then choose a wave
packet $\chi$ such that
\be
|(p_x^{\,2}-E)\chi|^2\leq \varepsilon/2\, . \label{chiasympt}
\ee
By making $\lambda$ sufficiently large we can make the ${\rm
O}(\lambda^{-1})$ corrections in \eqn{Hasympt} smaller than
$\varepsilon/2$. Combining \eqn{Hasympt} and \eqn{chiasympt} then 
shows that, with $\parallel \psi_\lambda\parallel = 1$,
\be
\parallel
(H-E)\psi_\lambda\parallel^2 \,\leq \varepsilon  
\ee
for any positive $E$ and $\varepsilon$. This proves that any
{\it nonnegative}
$E$ is a spectral value of the Hamiltonian $H$, so that
the spectrum is continuous. It is thus clear that the reason for the
continuity of the spectrum  is that wave functions can escape to
infinity along the valleys that have zero classical energy.
 
The above example exhibits the same qualitative features as the
supersymmetric quantum-mechanical models of section~1. Using the 
existence of the potential valleys extending 
to infinity and supersymmetry, it has been proven that also the 
Hamiltonian \eqn{hamiltonian} has a continuous spectrum starting 
at zero for any finite (compact) gauge group \cite{DWLN}. By 
contrast the spectrum of the corresponding bosonic theory is discrete 
\cite{LuescherS}. The mass operator of the supermembrane can 
be regarded as the $N\to\infty$ limit of the  supersymmetric 
Hamiltonian \eqn{hamiltonian} for the group $SU(N)$. Therefore, 
modulo unexpected subtleties associated with this  limit, the 
supermembrane mass spectrum is also continuous.

\section{Discrete states}
The result reported above indicates that the 
exact supermembrane is not a viable model for elementary 
particles. Still there remains the issue of finding discrete and 
possibly zero-energy eigenstates. Discrete eigenstates {\it 
within} the continuum are in principle possible, although such 
examples are not easy to construct and always
suffer from instabilities (as the discrete states
can ``decay" into the nearby continuum). The problem is somewhat 
simpler to address for the zero-energy states, where it boils 
down to the question of whether there exist 
{\it normalizable} solutions of $Q \,\psi = 0$. 

One might attempt to determine the nature of the ground state 
from the value of Witten index  
\cite{Witten}, which does not require an explicit derivation of 
the ground-state wave functions.
As is well known, if this index is nonzero there must be an unequal
number of zero-energy fermionic and bosonic states. At least some of
them must be annihilated by the supersymmetry charge.
Motivated by the connection of the supersymmetric quantum
mechanics with supersymmetric Yang-Mills theories, the Witten
index has been determined for $G=SU(2)$ in the so-called
ultralocal limit \cite{Smilga}. Usually one finds that the index 
is different from zero, with
the exception of the two-dimensional models (corresponding to a
supermembrane moving in a four-dimensional space-time), where
the index vanishes for odd-dimensional gauge groups $G$. In
\cite{PS} a ``twisted" version of the Witten index was proposed 
which was argued to be strictly positive. However, it
is known that, for a continuous spectrum without 
a gap, the very notion of the Witten index
will sensitively depend on the regularization that one employs 
\cite{contWI}. In any case, the issue cannot just be resolved 
by formal manipulations of certain functional integrals.
Rather than being positive the result is likely to be ill-defined!

Early attempts to determine directly whether or not the theory 
exhibits zero-energy ground states were described in \cite{DWHN}. 
Admittedly, the results of this work remained somewhat 
inconclusive, but all the evidence seemed to point 
in the direction of no normalizable  zero-energy states.  
Nevertheless, even for the simple two-dimensional model of the 
previous section, it is not known whether or not normalizable 
zero-energy states exist \cite{Kubek}. 

To illustrate some of the subtleties that one may encounter 
when analyzing this question, let us momentarily return to a 
variant of  
this two-dimensional model, where the supercharge depends on some 
superpotential $W(x,y)$,
\be
Q=\sigma_x\,p_x+\sigma_y\,p_y +\sigma_z\,W(x,y)\,. 
\label{Wsupercharge}
\ee
Possible valleys are associated with orbits consisting of points 
$(x,y)$ for which $W(x,y)$ 
vanishes. The presence of an arbitrary function $W$ offers a 
welcome feature, as it allows the study of the degenerate 
potentials as a limiting case of potentials that are better 
behaved and are therefore more amenable to rigorous analysis. 
However, here we simply  
restrict ourselves to the case that $W$ is rotationally invariant, so 
that it depends only on the radius $r$. This means that possible 
valleys take the form of circles in the $x$-$y$ plane. After 
passing to spherical coordinates, $r$  
and $\varphi$, one finds that the following two wave functions 
are annihilated by the supercharge \eqn{Wsupercharge},
\bea
\psi_\pm(r,\varphi)&=& {1\over \sqrt r}\,\exp\Big[\mp 
\int_{r_0}^r{\rm d}r^\prime\; W(r^\prime)\Big] \nonumber \\
&&\quad  \times  \pmatrix{{\rm e}^{-i\varphi/2}\cr 
\noalign{\vskip 4mm}
{\rm e}^{i(\varphi\pm\pi)/2}\cr}\,. \label{ground}
\eea
Depending on the behaviour of $W(r)$ at large $r$, one of these  
solutions can be square-integrable. A conspicuous 
feature is their behaviour at small distances. If $W(r)$ is regular 
near the origin, then the wave function exhibits a $1/\sqrt r$ 
singularity. One may wonder whether such a singularity, while 
leaving the wave function square-integrable, is 
acceptable. With such a singularity the 
kinetic-energy term in the Hamiltonian is not positive definite, 
although for \eqn{ground} the kinetic energy still cancels 
against the potential energy. Extra care is 
required when integrating by parts, where one may pick up  
boundary contributions at $r=0$.  

To analyze the situation in a little more detail, let us 
decompose the Hilbert space in eigenspaces of the angular-momentum 
operator. The corresponding wave functions can be written as 
\be
\psi(r,\varphi) = \pmatrix{{\rm e}^{-is\,\varphi}\, f_1(r)\cr 
\noalign{\vskip 4mm}
{\rm e}^{-i(s-1)\varphi}\, f_2(r)\cr}\,, 
\ee
where we expect that the relevant spin-values are given 
by $s=\ft12 (1+n)$ with $n$ an integer. One easily establishes 
that the supercharge \eqn{Wsupercharge} acts in subspaces of 
given spin. In each of these subspaces one can verify that the 
supercharge is self-adjoint (which is crucial for proving  
that the spectrum of the Hamiltonian is nonnegative!), provided 
the following condition is satisfied,
\be 
\lim_{r\downarrow 0}\, r\,(\phi , \s_x \psi ) = 0\,.
\label{hermQ}
\ee 
Here $\phi(r,\varphi)$ and $\psi(r,\varphi)$ are two 
two-component wave functions of the same spin (so that the 
expression in  \eqn{hermQ} is $\varphi$-independent) and $(\,,\,
)$ denotes the complex inner product for two-spinors. 
It turns out that precisely one of the wave functions \eqn{ground} can 
satisfy the above condition. Of course, this is no problem, as we 
already know that only one of them can be square-integrable. The 
same result is obtained when evaluating the condition required 
for the self-adjointness of the Hamiltonian. It reads 
\be
\lim_{r\downarrow 0}\, r\Big[ (\phi , \pa_r\psi ) - (\pa_r\phi, 
\psi )\Big]  = 0\,.           
\ee
Again we have to exclude one of the wave functions \eqn{ground} in 
order that the condition be satisfied. 

Hence we may conclude that certain mild singularities are acceptable 
in the wave function. This, however, has consequences for other 
potential eigenfunctions, so that a proper analysis is no longer 
restricted to a single wave function. Obviously, the presence of 
the singularities forms an extra complication in obtaining 
rigorous results for zero-energy states.     

Establishing the existence or nonexistence of zero-energy states 
for the supersymmetric models of section~1 with   
finite-dimensional groups, is even more difficult. 
The decomposition of the wave function in terms of the anticommuting 
coordinates gives rise to a multicomponent wave function. For 
the smallest case, namely that of the group $SU(2)$ in $d=2$ 
dimensions, the group-invariant states are characterized in terms 
of three-component    
wave functions. This theory was analyzed in \cite{DWHN}, where 
various arguments against the existence of zero-energy states were 
presented. Recently, a rigorous proof was presented that, 
indeed, normalizable zero-energy states do not exist for $SU(2)$ 
and $d=2$ \cite{hoppe2}.  

For $d=3$ there are also indications that no normalizable zero-energy 
states exist. Here 
we have coordinates $\vec X^A$ and complex two-component spinors 
$\theta^A$. The supercharges can be written as 
\bea
Q_\a &=& {1\over i}\, {\rm e}^W {\pa\over\pa \vec X{}^A}\, {\rm 
e}^{-W}  \,\Big( \vec\s\,\theta^A\Big)_{\!\a} \,,   \nonumber \\ 
Q^\dagger_a &=&  {1\over i}\, {\rm e}^{-W}{\pa\over \pa\vec 
X{}^A}\,{\rm e}^W \, \Big({\pa\over \pa \theta^{A}}\, 
\vec\s\Big)_{\!\a} \,,
\eea
and the superpotential $W$ reads 
\be
W= \ft16 f_{ABC}\, \varepsilon^{abc} \, X_a^A  X_b^BX_c^C\,.
\ee
Clearly $\exp[\pm W]$ is not normalizable as $W$ is odd in the 
coordinates. This directly precludes the existence of normalizable 
zero-energy states with zero or maximal fermion number. To 
determine whether the same result holds for states of 
intermediate fermion-number values, is not so easy, 
although it seems likely that the exponential factor will always 
cause a certain lack of normalizability. Unfortunately, more rigorous results  
regarding this question are still lacking. 
 
For other groups and in higher dimensions, an explicit evaluation 
of the resulting system of equations becomes untractable, 
because the number of components 
of the wave function increases exponentially with the  
spinor dimension times the group dimension.

\section{Interpretation of the continuum}

In the context of the elementary supermembrane, the continuity of 
the spectrum is clearly an undesirable result \cite{DWN}. 
However, one may turn things around and wonder whether there 
exists a context in  
which a continuous spectrum forms a more welcome feature. 
Obviously, such a spectrum would arise naturally when 
considering a many-particle Hilbert space 
of states associated with well-separated almost-free particles. 
Of course, the situation is rather subtle, as there should be no energy 
gap. The continuous spectrum starts at zero and possible bound 
states must be superimposed on this continuum.

There are objects that arise in the context of more 
complicated theories based on an infinite number of degrees of 
freedom, whose spectrum exhibits some of the 
characteristic features discussed above. The interesting point to 
note is that certain nontrivial properties of this spectrum may 
be reflected in the spectrum of a corresponding 
quantum-mechanical model. In this way these properties can thus 
be explored in the context of the conceptually more simple 
quantum-mechanical models based on a {\it finite} number of 
degrees of freedom.   

The above strategy has been applied successfully to the case of 
BPS monopoles and dyons of four-dimensional $N=4$ supersymmetric 
gauge theories. Static multi-monopole states of given energy and 
electric charge are contained in a 
multi-parameter family of solutions. The parameters that label 
these degenerate solutions are the collective coordinates, or 
moduli, and the space parametrized by these coordinates is called 
the moduli space. For multi-monopole solutions the collective 
coordinates are associated with the locations of the 
monopoles and their electric charge. In the supersymmetric case 
there will also be fermionic moduli. A detailed discussion of 
this is outside the scope of this lecture and we refer the reader 
to a recent set of lecture notes by Harvey \cite{Harvey}. The 
multi-monopole configurations can exist as static solutions, because 
the force between the static monopoles cancels.  
The repulsive Coulomb force cancels against the force mediated by 
the exchange of scalar particles. This cancelation can be 
understood from the fact that the Bogomol'nyi bound is saturated. 
It is known that the moduli space of BPS multi-monopole 
configurations must be  a hyper-K\"ahler space 
\cite{AtiyahH}. 

One may also consider configurations of slowly moving monopoles 
by allowing the collective coordinates to depend on  
time. The dynamics of the multi-monopole solutions can then be 
described in terms of a relatively simple quantum-mechanical 
model defined on a $4n$-dimensional hyper-K\"ahler space, where 
$n$ denotes the number of monopoles. In $N=4$ Yang-Mills theory 
the BPS monopoles break half of the number of supersymmetries, so 
that the moduli space is extended with $8n$ fermionic collective 
coordinates, which are grouped into $4n$ real spinor doublets 
related by supersymmetry to the $4n$  
bosonic moduli. After extracting the centre-of-mass collective 
coordinates corresponding to position and total charge, 
one is thus left with supersymmetric quantum mechanics based on a 
$(4n-4$)-dimensional K\"ahler space \cite{GauntlettB}. The model 
has four independent supersymmetries parametrized by real 
anticomuting doublet parameters, which are related to the eight 
supersymmetries that were left invariant in the underlying field 
theory by the BPS configurations.  

From a conjectured property of the four-dimensional gauge theory, 
namely $S$  
duality, one deduces that there should exist a bound state that 
satifies the Bogomol'nyi bound with electric charge $q$ and 
magnetic charge $n$, for each $q$ and $n$ relatively prime. This 
implies that the corresponding ($4n-4$)-dimensional quantum 
mechanics model must have a supersymmetric ground state. For this  
type of models a supersymmetric state requires the existence of a 
certain harmonic form on the hyper-K\"ahler space \cite{witten2}. 
The above arguments were presented by Sen \cite{Sen}, who then 
demonstrated, using existing results for the two-monopole moduli 
space \cite{GibbonsM}, that a corresponding harmonic form of the  
required characteristics does exist. A more general proof for an  
arbitrary number of monopoles was presented in \cite{Porrati}. 

Hence from a study of supersymmetric systems based on a finite 
number of degrees of freedom, one can verify the implications of 
a nonperturbative field-theoretic phenomenon, such as $S$ duality. 
However, we should point out that the models that are relevant for 
the multi-monopole states belong to a different class than the models 
of section~1. Unlike the latter they are not defined in terms of 
a Lie group and neither are they subject to a  
Gauss constraint. They  can be obtained upon dimensional 
reduction from supersymmetric sigma models and not from 
supersymmetric gauge theories. 

The models of section~1 have another role to play. In string 
theory strings can end on certain defects by means of  
Dirichlet boundary conditions. These defects are therefore 
called D-branes (for further references, see \cite{Dbranes}). 
They can have a certain $p$-dimensional spatial extension,  
and carry Ramond-Ramond charges \cite{Polchinski}. 
These D-branes play an important role in the nonperturbative 
behaviour of string theory. Here we concentrate on 
the  D0-branes (or Dirichlet particles). 

It is not so difficult to determine the effective short-distance 
description for D-branes \cite{boundst}. As the strings must be 
attached to the  $p$-dimensional branes, we are dealing with open 
strings whose endpoints are attached to a  
$p$-dimensional subspace. At short distances, the interactions 
caused by these open strings are determined by the massless 
states of the open string, which constitute the ten-dimensional 
Yang-Mills supermultiplet, propagating in a ($p+1$)-dimensional 
spacetime. Because the endpoints of open strings carry Chan-Paton 
factors the effective  
short-distance behaviour of $n$ D-branes is described in terms 
of an $U(n)$ ten-dimensional supersymmetric gauge theory reduced 
to the $(p+1$)-dimensional world volume of the D-brane. The 
$U(1)$ subgroup describes the centre-of-mass motion of the $n$ 
D-branes. 

In the type-IIA superstring one is dealing with Dirichlet 
particles moving in nine dimensions. As the world volume is 
one-dimensional ($p=0$), the short-distance interactions between 
these particle is thus described by the model of section~1 with 
gauge group $U(n)$ and $d=9$. The continuous spectrum without 
gap is natural here, as it is known that, for static D-branes, the 
Ramond-Ramond repulsion cancels against the gravitational and 
dilaton atraction, a similar phenomenon as for BPS 
monopoles \cite{Polchinski}. With this gauge group the  
coordinates can be described in terms of $n\times n$ hermitean 
matrices. The valley configurations correspond to the situation 
where these matrices can be diagonalized simultanously. Hence the 
coordinates of the model are restricted to nine 
diagonal matrices whose eigenvalues define the positions of $n$ 
D-particles. As soon as one or several of these particles coincide  
then the $[U(1)]^n$ symmetry that is left invariant in the 
valley, will be enhanced to a nonabelian 
subgroup of $U(n)$. Clearly the degrees of freedom of the model 
are more than just the D-particle coordinates. The additional 
degrees of freedom correspond to the strings stretching between 
the D-particles. Note also that the model naturally 
incorporates configurations corresponding to widely separated 
clusters of D-particles, each of which can be described by a 
supersymmetric quantum mechanics model based on $U(k)$ subgroups 
of $U(n)$. When all the D-particles move further apart this 
corresponds to configurations deeper and deeper into the 
potential valleys.  

The D-particles thus define a new perspective on the models 
introduced in section~1. Their dynamics can 
be studied in these models and we refer to \cite{Dpart,DKPS} for 
work along these lines. We should also mention that, depending on 
the kind  
of branes one is dealing with, other variants of supersymmetric 
quantum mechanics may be relevant. These can be obtained from  
the zero-volume limit of supersymmetric gauge theories coupled to 
matter. For instance, one such model was used in \cite{DKPS} to 
study the  
interaction between a 0-brane and a 4-brane. The two-dimensional 
model of section~4 may be regarded as a truncation of this 
nine-dimensional system. 

From the quantum mechanics based on a finite number of degrees of 
freedom one may again hope to learn about the properties of 
D-branes. Some of these properties are again gouverned by 
duality arguments, in this case related to M-theory. This theory 
is conjectured to be the unified theory that encompasses all 
string theories (for a recent review, see \cite{Mtheory}). It is 
eleven-dimensional in origin, but its   
structure is still largely unknown. String theories themselves do 
not live in eleven dimensions and should be viewed as 
perturbative realizations of M-theory. The  small coupling 
is thus tied to the small size of certain compactified 
dimensions. At large   
distances M-theory should obviously be described  
by eleven-dimensional supergravity. When the size of the 
compactification is shrunk to zero, one describes the massless 
excitations of the corresponding string theory. The simplest such 
compactification, namely where one dimension is compactified on a 
circle, leads to IIA string theory. The size of the compactified  
coordinate, denoted by $R_{11}$ is identified as the ratio of the 
string coupling constant $g_{\rm s}$ and the string mass scale $M_{\rm 
s}$. By the standard Kaluza-Klein arguments, $R_{11}$ is also related to 
the Planck masses in eleven and ten dimensions, so that we obtain  
\be
R_{11}= {g_{\rm s}\over M_{\rm s}}= {[M^{\rm Planck}_{10}]^8 
\over  [M^{\rm Planck}_{11}]^9}\,.   
\ee
Furthermore we have the familiar ten-dimensional string 
relation between the Planck constant and the string scale, 
$M_{\rm s}^8 = g^2_{\rm s} \,[M^{\rm Planck}_{10}]^{8}$. 
Combining these equations leads to $M_{\rm s} = g^{1/3} M^{\rm 
Planck}_{11}$ as well as to $R_{11}= g_{\rm s}^{2/3}/M^{\rm 
Planck}_{11}$.

By various compactifications one can obtain the different types of 
string theories. In this way, the existence of M-theory explains 
the string-string dualities that have  
been discovered in the last few years. These dualities were found 
both in  the effective low-energy field theories based on 
different string vacua and by including solitonic $p$-branes into 
the various string theories and comparing their spectra (see e.g. 
\cite{Townsend,witten3}). 

The D-particles of IIA string theory can now be regarded as 
Kaluza-Klein states emerging in the compactification of 
eleven-dimensional supergravity on $S^1$. The Kaluza-Klein states 
carry charges which are multiples of $1/R_{11}$ and the 
Kaluza-Klein photon couples to them. However, viewed from a 
string perspective, the Kaluza-Klein photon corresponds to one of 
the Ramond-Ramond fields of IIA string theory, so that the 
massive Kaluza-Klein states carry Ramond-Ramond charge. 
Consequently some of them can be identified with the 
D-particles. On the other hand, the IIA supergravity theory itself has 
solitonic solutions, extremal black holes, which are charged with 
respect to the Ramond-Ramond photon \cite{HorStrom}. These 
particles are BPS states and leave only half the ten-dimensional 
supersymmetries invariant, precisely the same number as is left 
invariant by the massive Kaluza-Klein states! Hence they can be 
identified as D-particles \cite{Townsend}.

Just as for the BPS monopoles, discussed earlier, one expects 
bound states for D-particles, precisely at threshold. Their 
existence can be inferred from the Kaluza-Klein spectrum of 
eleven-dimensional supergravity. In this 
way one may be able to verify some of the implications of duality with 
M-theory in the context of the models introduced in section~1. 
However, as we discussed in the previous section, to  
prove or disprove the existence of normalizable states within the 
continuum, is still a formidable task. 

In summary, the situation is quite intriguing. On the one hand, 
supermembranes are just models in quantum mechanics associated 
with $U(N)$ in the large-$N$ limit. These supermembranes can live 
in a background of eleven-dimensional supergravity, which in itself 
is a very nontrivial property, but 
they are not viable as an elementary supermembrane in view of 
their continuous mass spectrum. On the other hand, the 
eleven-dimensional membrane provides, upon suitable compactification  
and in certain truncations, the various 
string excitations, including the solitonic ones. At the same 
time, the $N\to \infty$ quantum mechanics that describes the 
supermembrane, can be regarded as a theory of 
infinitely many D-particles. In a Kaluza-Klein perspective, these 
are just the massive Kaluza-Klein states of eleven-dimensional 
supergravity, which, when combined 
with the ten-dimensional IIA supergravity states, constitute 
eleven-dimensional supergravity compactified on a circle of finite 
radius. Apparently we are dealing with many faces of a single 
theory. In fact, one could turn the argument around and argue 
that it is the quantum-mechanical models that are fundamental. 
Recently this has led to  
the conjecture that the $N\to \infty$ limit is in fact a 
representation of uncompactified M-theory \cite{BFSS}. 
Irrespective of whether this is the right approach or not, these 
and related questions will eventually be clarified in the context 
of the elusive M-theory.

\vspace{5mm}
\par
\noindent{ \bf Acknowledgements}\vspace{0.3cm}
\par
A substantial part of this work is based on results reported in 
\cite{DWLN,DWHN,DWMN,DWN}. The work is supported by the European 
Commission TMR programme ERBFMRX-CT96-0045.  I am grateful to H. 
Nicolai and E. Rabinovici for valuable discussions.



\begin{thebibliography}{99}                                     

\bibitem{CH} M. Claudson and M.B. Halpern, Nucl. Phys. B250 (1985) 
689; 
R. Flume, Ann. Phys. {164} (1985) 189; 
M. Baake, P. Reinicke, and V. Rittenberg, J. Math.
Phys. {26.} (1985) 1070. 
\bibitem{LuescherS} M. L\"uscher, Nucl. Phys. B219 (1983) 233; 
B. Simon, Ann. Phys. 146 (1983) 209.
\bibitem{DWLN}
 B. de Wit, M. L\"uscher and H. Nicolai, Nucl. Phys. B320 (1989) 
135.
\bibitem{BST} E. Bergshoeff, E. Sezgin and P.K. Townsend, Phys. 
Lett. 189B (1987) 75.
\bibitem{BTD}
 E. Bergshoeff, E. Sezgin and P.K. Townsend,
 Ann. Phys. 185 (1988) 330; 
 P.K. Townsend, in "Superstrings '88", proc. of the Trieste Spring
School, eds. M.B. Green, M.T. Grisaru, R. Iengo and A. Strominger
(World Scient., 1989); 
 M.J. Duff, Class. Quantum Grav. 5 (1988) 189.
\bibitem{AETW} A. Ach\'ucarro, J.M. Evans, P.K. Townsend and D. L.
Wiltshire, Phys. Lett. 198B (1987) 441.
\bibitem{DWHN}
 B. de Wit, J. Hoppe and H. Nicolai, Nucl. Phys. B305 [FS23]
(1988) 545.
\bibitem{Goldstone} J. Goldstone, unpublished.
\bibitem{Hoppe} J. Hoppe, in proc. Int. Workshop on
Constraint's Theory and Relativistic Dynamics, eds. G. Longhi and
L. Lusanna (World Scient., 1987).
\bibitem{BSTa} E.A. Bergshoeff, E. Sezgin and Y. Tanii, Nucl. 
Phys. B298 (1988) 187.
\bibitem{DWMN} B. de Wit, U. Marquard and H. Nicolai, Commun. 
Math. Phys. 128 (1990) 39. 
\bibitem{DWN} B. de Wit and H. Nicolai, in proc. Trieste Conference on 
Supermembranes and Physics in $2+1$ dimensions, p. 196, eds. M.J. Duff, 
C.N. Pope and E. Sezgin (World Scient., 1990). 
\bibitem{Witten} E. Witten, Nucl. Phys. {B188} (1981) 513.
\bibitem{Smilga} A.V. Smilga, Yad. Fiz. 42(1985) 728; 43 (1986) 215;
Nucl. Phys. B266 (1986) 45.
\bibitem{PS} C.N. Pope and K.S. Stelle, Class. Quantum Grav. 5
(1988) L161.
\bibitem{contWI}
A.J. Niemi and L.C.R. Wijewardhana, Phys. Lett. 138B (1984) 
389; 
R. Blanckenbeckler and D. Boyanosvsky, 
Phys. Rev D30 (1984) 1821; D31 (1985)2089; 
D. Boll\'e, F. Gesztesy, H. Grosse, W. Schweiger and B. Simon,
J. Math. Phys. 28 (1987) 1512.
\bibitem{hoppe2} J. Fr\"ohlich and J. Hoppe, {\it On zero-mass 
ground states in super-membrane matrix models}, hep-th/9701119. 
\bibitem{Kubek}
A. Kubek, Diplomarbeit (unpublished).
\bibitem{Harvey} J. Harvey, {\it Magnetic Monopoles, Duality and 
Supersymmetry}, hep-th/9603086.
\bibitem{AtiyahH} M. Atiyah and N. Hitchin, Phys. Lett 107A 
(1985) 21; 
{\it The Geometry and Dynamics of Magnetic Monopoles}, Princeton 
Univ. Press, 1988. 
\bibitem{GauntlettB} J.P. Gauntlett, Nucl. Phys. B400 (1993) 103, 
B411 (1994) 443;  
J. Blum, Phys. Lett. B333 (1994) 92.
\bibitem{witten2} E. Witten, Nucl. Phys. B202 (1982) 253.
\bibitem{Sen} A. Sen, Phys. Lett. B329 (1994) 217.
\bibitem{GibbonsM} 
G.W. Gibbons and N.S. Manton, Nucl. Phys. B274 (1986) 183, Phys. 
Lett. B356 (1995) 32. 
\bibitem{Porrati}  M. Porrati, Phys. Lett. B377 (1996) 76. 
\bibitem{Dbranes} J. Polchinski, S. Chaudhuri and C.V. Johnson, 
{\it Notes on D-branes}, hep-th/9602052; J. Polchinski, {\it TASI 
Lectures on D-branes}, hep-th/9611050; C. Bachas, {\it (Half) A 
lecture on D-branes}, lectures given at the Workshop on Gauge 
Theories, Applied Supersymmetry and Quantum Gravity, London, 1996 
(hep-th/9701019).   
\bibitem{Polchinski} J. Polchinski, Phys. Rev. Lett. 75 (1995) 
4724 (hep-th/9510017).
\bibitem{boundst} E. Witten, Nucl. Phys. B460 (1996) 335 
(hep-th/9510135).
\bibitem{Dpart}
U. Danielson, G. Ferretti and B. Sundborg, Int. J. Mod. Phys. A11 
(1996) 5463 (hep-th/9603081); 
D. Kabat and P. Pouliot, Phys. Rev. Lett. 77 (1996) 1004 
(hep-th/9603127).
\bibitem{DKPS}
M.R. Douglas, D. Kabat, P. Pouliot and S.H. Shenker, {\it 
D-branes and short distances in string theory}, hep-th/9608024.
\bibitem{Mtheory} P.K. Townsend, {\it Four lectures on M-theory}, 
lectures given at the 1996 ICTP Summer School in High Energy 
Physics and Cosmology, Trieste (hep-th/9612121).
\bibitem{HorStrom} G. Horowitz and A. Strominger, Nucl. Phys. 
B360 (1991) 197. 
\bibitem{Townsend} P.K. Townsend, Phys. Lett. B350 (1995) 184 
(hep-th/9501068), Phys. Lett. B373 (1996) 184 (hep-th/9512062).  
\bibitem{witten3} E. Witten, Nucl. Phys. B443 (1995) 85 
(hep-th/9503124). 
\bibitem{BFSS}
T. Banks, W. Fischler, S.H. Shenker and L. Susskind, {\it 
M-theory as a matrix model: a conjecture}, hep-th/9610043. 



\end{thebibliography}
\end{document}